\documentstyle[preprint,aps,prl]{revtex}
\begin{document}
\draft
\title{DISCRETE BREATHERS AND
ENERGY LOCALIZATION IN NONLINEAR LATTICES}
\author{Thierry Dauxois and Michel Peyrard}
\address{
Laboratoire de Physique de l'Ecole Normale Sup\'erieure de Lyon,
CNRS URA 1325,
46 all\'ee d'Italie, 69007 Lyon, France.}
\date{\today}
\maketitle
\begin{abstract}
We discuss the process by which energy,
initially evenly distributed in a
nonlinear lattice, can localize itself
into large amplitude excitations. We
show that, the standard modulational
instability mechanism, which can
initiate the process by the formation
of small amplitude breathers, is
completed efficiently, in the presence
 of discreteness,
by energy exchange mechanisms between
the nonlinear
excitations which favor systematically
the growth of the larger excitations.
The process is however self regulated
 because the large amplitude
excitations are finally trapped by
the Peierls-Nabarro
potential.
\end{abstract}

\pacs{PACS numbers: 63.10+a, 3.40 Kf, 46.10+z}
\narrowtext

\section{INTRODUCTION}
\label{sec:intro}

Many physical phenomena involve some
localization of energy in space.
The formation of vortices in hydrodynamics, self
focusing in optics or plasmas,
the formation of dislocations in solids
 under stress, self trapping
of energy in proteins, are well known examples.
Following the original work by Anderson
 \cite{ANDERSON}
disorder-induced localization has been
 widely studied, but, more recently,
attention was attracted on the possibility
to localize energy in an
homogeneous system due to nonlinear
effects. the process can become
dramatic when it leads to collapse
in a plasma \cite{ZAKHAROV}.
In this paper we are
interested in the process by which
energy evenly
distributed in such a system
can concentrate itself spontaneously
into spatially localized
nonlinear excitations.
In some cases this evolution can lead
to the formation of topological
solitonlike excitations such as
dislocations or ferroelectric or
ferromagnetic domain walls. However,
since there is an energy threshold
for the creation of topological solitons, the first step
of the evolution is
the formation of breathers or envelope modes; we shall
therefore focus our attention on such modes.

Nonlinear energy localization in
continuous media has been extensively
investigated since Benjamin and Feir
 \cite{BENJAMIN} discovered
the modulational instability of  Stokes
 waves in fluids, but
very little has been done in lattices although it
would be of wide interest for solids
 or macromolecules.
We want to point out
here that, in a discrete lattice,
nonlinear energy localization is very
different from its counterpart in a continuum medium.
In particular, we show
that, besides the familiar mechanism
 of modulational instability, which is
itself strongly modified by discreteness
effects, there is an additional
channel for energy concentration, which
is specific to lattices, but is
not sensitive to the details of the
nonlinear lattice model which is
considered. Therefore it appears as a
very general process leading
to localization of energy in a lattice.

The first step toward the creation of
localized excitations can be
achieved through modulational instability
 which exists in a lattice as well
as in a continuum medium, although discreteness can
drastically change the conditions
for instability  \cite{KIVSHAR}
(e.g., at
small wave numbers a nonlinear carrier
 wave is unstable to {\it all}
possible
modulations of its amplitude as soon
as the wave amplitude exceeds a certain
threshold). However the maximum energy
of the breathers created by
modulational instability is bounded
because each breather
collects the energy
of the initial wave over the modulation
length $\lambda$ so that its energy
cannot exceed $E_{\text {max}} =
\lambda \, e$ where $e$
is the energy density of the
plane wave. Consequently, although
 modulational instability can lead to a
strong increase in energy {\it density}
in some parts of the system, it
cannot create breathers with a {\it total}
energy exceeding $E_{\text
{max}}$. For a given initial energy density,
one can however go beyond this
limit if one excitation can collect
the energy of several breathers
created by modulational instability.
Such a mechanism is not observed
in a continuum medium because there
the breathers generated by modulational instability
are well approximated by solitons of
the Nonlinear Schr\"odinger (NLS)
equation which can pass through each
other without exchanging energy.
On the contrary, when discreteness effects
are present, the
energy of each excitation is {\it not}
conserved in collisions, and, the
important point is that {\it the exchange
tends to favor
the growth of the larger excitation}. In
order to analyze the growth of the
breathers in a lattice, we must therefore
examine three of their properties:
i) their stability, ii) their ability to
move in the lattice, iii) the nature
of their interactions.

In order to discuss these points
quantitatively, let us, in a first step,
examine a specific model.  We consider
a chain of harmonically coupled
particles situated at positions $u_n$
 and submitted to the substrate potential
\begin{equation}
\label{eq:epotentiel}
V(u_n)=\omega_d^2\left({u_n^2\over 2}-{u_n^3\over3}\right)\; ,
\end{equation}
where $\omega_d^2$ is a parameter
 which measures the amplitude of the
substrate potential, and therefore
controls discreteness.
We will be interested in motions
inside the potential well ($u<1$).
This potential can
be viewed as a medium amplitude expansion of
any asymmetric potential around
a minimum. It can for instance represent the
expansion of a Morse potential
in a nonlinear model for DNA denaturation
\cite{DAUXOIS} or the expression
around a minimum of the well known $\phi^4$
potential \cite{CAMPBELL}.
The hamiltonian of the model is
\begin{equation}
\label{eq:hamiltonian}
 H =\sum_{n}^{}\Biggl[ {1\over 2} \dot u^{2}_n
 +{1\over 2} (u_n-u_{n-1})^{2} +V(u_n) \Biggr]\; .
\end{equation}
The existence and stability of breathers
 in nonlinear Klein-Gordon models
has been the subject of many investigations
\cite{CAMPBELL} and is not yet
completely understood. However, we have
shown that, provided that
discreteness is strong enough, extremely
stable large amplitude breathers
can exist in such a model  \cite{DAUXOIS}.
They can be obtained
with the Green's function
method introduced by Sievers and Takeno
\cite{SIEVERS} for intrinsic
localized modes in lattices with anharmonic
 coupling. The role of
discreteness to stabilize the breathers
 can be understood if one starts from
the ``anti-integrable'' limit where the
on-site nonlinear oscillators are
decoupled and then turns on a coupling
which remains weak with
respect to the
on-site potential \cite{AUBRY}. Thus
discrete breathers are sufficiently
stable to have a long lifetime which
gives them sufficient
time to interact, provided that they
can move in the lattice. This point is
not as trivial as it might seen.

\section{PEIERLS-NABARRO BARRIER FOR A BREATHER}
\label{sec:PNBA}

The trapping effect of the discreteness
 is well known for topological
solitonlike excitations and has been
 extensively investigated in the context
of dislocation theory \cite{SEEGER}.
 In a lattice, a kink cannot move
freely. The minimum energy barrier
 which must be overcome to translate
the kink by one lattice period is
known as the Peierls-Nabarro (PN) barrier,
$E_{\text {PN}}$. It can be calculated
by evaluating the energy of a static
kink as a function of its position in
 the lattice.
For the various models which have been
 investigated, two extremal values are
generally obtained when the kink is
exactly situated on a lattice site
(centered solution) or when it is in
the middle between two sites
(non-centered solution).

For a discrete breather  very little is known,
 although the PN barrier has
been shown to exist \cite{CAMPBELL}. One of the
difficulties is that the breather is a
 two-parameter solution. While for a
kink, the PN barrier depends only on
 discreteness, {\it i.e.} on the model
parameters, for a breather it depends
also upon its amplitude (or
frequency). This amplitude dependence
 is crucial for our analysis because we
are interested in the growth of breathers.
As they increase in amplitude,
the PN barrier that they feel changes.
The definition of the Peierls
barrier itself is not as simple for
 a breather as for a kink. In
principle, its value can be obtained by
 monitoring the breather as it
is translated along one lattice constant.
 While for a kink the path
followed by the particles in the multidimensional
phase space of the system
can be obtained by minimizing the energy while the
position of the central particle is
 constrained in all intermediate
states, in the case of the breather,
the path in the phase space is  not
a minimum energy path but a succession
of saddle points. The
energy of a kink which  is exactly
centered on a site or in the middle
between two sites is defined without
 ambiguity. For a breather with
a given frequency when it is centered
 on a site, there is no obvious
constraint which imposes that it
should have the same frequency when it
is situated in the middle between
two sites. We have used, as a working
definition of the PN barrier for
a breather the difference between
the energies of a centered and a
 non-centered breather {\it with the
same frequency}. This definition
 gives results which
agree with the observations of the
breather motion made by molecular
dynamics simulations, but the
notion of PN barrier for a breather
will require further analysis.

\subsection{Large amplitude breathers}
\label{subsec:Lab}
 To calculate the PN-barrier,
we have to compare the energy between
 two cases: the breathers is
centered on a particle or between
 particles. In the previous paper
\cite{DAUXOIS}, we focused our study
in the first case,
but we can easily extend the method
to the second one.
The procedure is the following~:
 we look for stationary-mode solutions
by putting
\begin{equation}
u_n=\sum_{i=0}^{\infty}\phi_n^i\ \cos(i\omega_b t)\; ,
\label{eansatz}
\end{equation}
where $\omega_b$ is the eigenfrequency
of the breather and $\phi_n^i$
are time independent amplitude oh the
 $i^{{th}}$ mode.
 Inserting the ansatz
(\ref{eansatz}) in the dimensionless equation of motion,
 we set the coefficients of
$\cos(i\omega_b t)$ equal to each other,
 retaining only the first
 three terms. We obtain~:
\begin{mathletters}
\label{eq:egreen}
\begin{eqnarray}
\omega_d^2\ \phi_n^0-\bigl[\phi_{n+1}^0+
\phi&&_{n-1}^0-2
\phi_n^0\bigr]\nonumber\\
&&=\omega_d^2\ \bigl[{\phi_n^0}^2+
{{\phi_n^1}^2+{\phi_n^2}^2
\over 2}\bigr]\label{eq:egreena}
\end{eqnarray}
\begin{eqnarray}
(\omega_d^2-\omega_b^2)\ \phi_n^1-\bigl[
\phi_{n+1}^1+\phi&&_{n-1}
^1-2\phi_n^1\bigr]\nonumber\\
&&=\omega_d^2\ \bigl[2\phi_n^0+\phi_n^2
\bigr]\ \phi_n^1  \label{eq:egreenb}
\end{eqnarray}
\begin{eqnarray}
(\omega_d^2-4\omega_b^2)\ \phi_n^2-
\bigl[\phi_{n+1}^2+\phi&&_{n-1}
^2-2\phi_n^2\bigr]\nonumber\\
&&=\omega_d^2\ \bigl[2\phi_n^0\phi_n^2+{{
\phi_n^1}^2\over 2}\bigr]\label{eq:egreenc}
\end{eqnarray}
\end{mathletters}
\narrowtext
Then, invoking the Green's functions
for the linear left-hand sides,
 we get a set of simultaneous
nonlinear eigenvalue equations
determining the eigenfrequency
$\omega_b$ and the eigenfunctions $\phi_n^i$ ~:
\begin{mathletters}
\label{eq:egreenbis}
\begin{equation}
\phi_n^0=\sum_{m}^{} G(n-m,0)\
\bigl[{\phi_m^0}^2+
{{\phi_m^1}^2+{\phi_m^2}^2\over 2}
\bigr]\label{eq:egreenbisa}
\end{equation}
\begin{equation}
\phi_n^1=\sum_{m}^{} G(n-m,\omega_b)\ \bigl[2
\phi_m^0+\phi_m^2\bigr]\ \phi_m^1
\label{eq:egreenbisb}
\end{equation}
\begin{equation}
\phi_n^2=\sum_{m}^{} G(n-m,2\omega_b)\ \bigl[2
\phi_m^0
\phi_m^2+{{\phi_m^1}^2\over 2}\bigr]
\label{eq:egreenbisc}
\end{equation}
\end{mathletters}
\narrowtext
where the Lattice Green's functions
have the following expression~:
\begin{equation}
G(n,\omega_b)={\omega_d^2\over N}
\sum_{q}^{} {e^{iqn}
\over \omega_d^2-\omega_b^2+2[1-\cos(q)]}\; .
\label{efunctionGReen}
\end{equation}

For solving this system, the procedure requires
more care than in the centered case\cite{DAUXOIS},
to avoid the problem of instability of this mode.
 Indeed,
since the position at the top of the
PN barrier is intrinsically unstable,
regardless of other possible causes
 of instability, even starting
with a symmetrical initial condition,
 the results show that in all cases
the breather moves so that the center reaches
  the bottom of the well.
{\it i.e.} the solution converges
toward the more stable breather.
 To prevent this tendency we chose
 to impose the symmetry
 and calculate the solution for only
 a half of the chain,
 the second half being know by symmetry~:
thus the position of the breather
is fixed. Self-consistently solved,
the system (\ref{eq:egreenbis})
 give us the values
of the breather's frequency and of
the amplitude of the differents sites.

In Fig.~\ref{fig:omegaamplitude}(a), we plot
 the amplitude versus the frequency
for the two modes. At high frequency
({\it i.e.} low amplitude),
 the two curves are very close to
each other. As one might expect
(see Fig.~\ref{fig:profilsolu}),
the amplitude of
 the mode centered on a particle
is larger than when the mode is
 centered between particles.
 A comparison of the energies in
 the two cases reveals a
great difference as shown in
Fig.~\ref{fig:omegaamplitude}(b).
 The solution centered
on a particle has a much lower energy.
Indeed, as the discreteness effects
are important, the substrate energy
is the dominant contribution to the
total energy. When the breather
is centered between particles, two of them
 participate mainly to the excitation,
 giving rise to a substantial increase
 of the energy, in comparison with the previous case,
 where only one particle has a big amplitude.

Figure \ref{fig:instabilite} illustrates the simulation
of the dynamics of the breather with
 a decentered solution as an initial condition.
The numerical scheme for solving the nonlinear equations
of motion
is a fourth order Runge-Kutta method
of a lattice with
 typically 256 atoms and periodic
boundary conditions.
Starting with a breather centered
between the sites
24 and 25, and with a frequency
 $\omega_b=0.93\ \omega_d$, the pictures
\ref{fig:instabilite}(a), \ref{fig:instabilite}(b)
 and  \ref{fig:instabilite}(c) show
the envelop of the oscillations
of the particles 24, 26 and
25. It is clear that after about
 70 breather's oscillations,
 the excitation moves to be centered
on  particle 25.
 Although we start with a perfect
symmetrical initial condition
 centered between  particle,
 because of its intrinsic
instability, some unavoidable
numerical errors have moved
 the breather down the PN
 barrier. As the initial condition
is clearly not the  exact solution
at the bottom of the well, a
 phenomenon of modulation appears:
 it is a consequence\cite{DAUXOIS}
of the combination
of the breather's frequency ($0.88\
  \omega_d$ after the displacement
of the breather) with the
 mode situated exactly at the
 bottom of the phonon band,
 which cannot be radiated away
because of its zero group velocity.

 Figs.~\ref{fig:instabilite}(a) and \ref{fig:instabilite}(b)
 show the oscillations of the two nearest
neighbors of the center.
It is clear that, after the translation
 of the center, they have a
similar evolution,
except that a new modulation effect
is present: it is  due to the
 combination of the
former frequency with the frequency of
 the oscillation in the well of the PN barrier.
Furthermore the two particles
are not in phase, because
when one starts with a breather,
the shape mode (the
derivative with respect to the position
of the  breather's center) is odd:
 so, with the center on a particle,
 the shape mode is such that
 the central site does not move,
 but the two neighboring sites
 are 90 degrees out of phase.

As the spontaneous evolution of
a decentered breather with a
frequency $\omega$ gives a
centered breather with a lower
frequency, we can not calculate
 the PN barrier from such a simulation.
Although the energy of the system
is conserved  during the integration
 of the equations,
it does not stay localized in the
 excitation since the movement
of the breather generates a strong
 radiation of phonons.
The Peierls barrier can be
 obtained from a calculation
of the energy of the solutions
given by the Green's
functions method, independently
in the two cases.
Figure \ref{fig:omegaamplitude}(b)
shows that  the PN barrier is very high;
as it is an increasing function
 of the amplitude of the breather
(see  Fig.~\ref{fig:omegaamplitude}(a)),
the barrier is a decreasing function
of the amplitude. A small amplitude
 breather will propagate easily
the chain along, whereas the large
 amplitude ones will be trapped
on a site because of the additional
potential due to the discreteness.

\subsection{Approximate analytical
expression and the PN barrier}
\label{subsec:Ae}

The Green's function method provides a
very accurate expression for
the discrete breather modes,
but the solution is known only numerically.
When the breathers are highly
localized, it is possible to derive
an approximate analytical solution
for the frequency
of the mode versus their amplitude
in the two cases.
 First, we consider the centered case
 where the mode is on a particle
which we call
$n=0$. As the mode is highly localized,
we assume $|u_n|\ll |u_1|$ for $|n|> 1$.
 We seek an approximate solution of the dimensionless
 equation of motion, by looking
for a solution which
is localized over only 3 sites,
 putting\cite{KISHVAR}:
\begin{equation}
 u_0= A+B\cos(\omega_b t)
\label{ansatzun}
\end{equation}
\begin{equation}
 u_1=u_{-1}= C+D\cos(\omega_b t)\; .
\label{ansatzdeux}
\end{equation}
\begin{equation}
 u_l=0\; for\; |l|\ge1\; .\label{ansatztrois}
\end{equation}
 The dc parts of the ansatz are positive
because of the
asymmetry of the potential.
 We insert this ansatz in the dimensionless equation of motion, and set
the coefficients of $\cos(\omega_b t)$
and the constant term
equal to each
other. We obtain for $n=0$:
\begin{equation}
0=2(C-A)-\omega_d^2(A-{B^2\over2}-A^2)
\label{enegalzeroconst}
\end{equation}

\begin{equation}
 -\omega_b^2 B=2(D-B)-\omega_d^2(B-2AB)
\label{enegalzeroomeg}
\end{equation}
For $n=1$, it yields:
\begin{equation}
0=A-2C-\omega_d^2(C-{D^2\over2}-C^2)
\label{enegalunconst}
\end{equation}

\begin{equation}
-\omega_b^2D=B-2D-\omega_d^2(D-2CD)
\label{enegalunomeg}
\end{equation}

As  the excitation is rapidly decreasing,
 we can estimate that $A\gg C$;
 equation (\ref{enegalzeroconst}) gives then:
\begin{equation}
B=\sqrt{2A\left({2\over \omega_d^2}
+1-A\right)}\; ,\label{egrandB}
\end{equation}
whereas (\ref{enegalzeroomeg}) and
(\ref{enegalunomeg}) gives:
\begin{equation}
\omega_b^2=2\left(1-{D\over B}\right)
+\omega_d^2(1-2A) 
\end{equation}
\begin{equation}
=2-{B\over D} +\omega_d^2(1-2C)\label{eomegadeux}
\end{equation}

Neglecting $C$ in (\ref{eomegadeux}),
 we obtain from these two equations:
\begin{equation}
\left({D\over B}\right)^2+\omega_d^2A\
\left({D\over B}\right)-{1\over 2}=0\; .
\end{equation}
Then, we get:
\begin{equation}
{D\over B}={-A\omega_d^2\pm
\sqrt{(A\omega_d^2)^2+2}\over 2}\; .
\label{eDsurB}
\end{equation}
As we are interested in breather modes,
the particles should oscillate in phase:
the ratio ${D/B}$ must be
positive and that's why we keep only the plus sign.
Equation (\ref{eomegadeux}) gives then:
\begin{equation}
{\omega_b^2\over \omega_d^2}={2\over
 \omega_d^2}+1-A-\sqrt{{2\over
\omega_d^4}+A^2}\label{eomegaamplapprox}
\end{equation}
Consider now the case where the
 center of the excitation is between
two particules (in our notation,
between the site (-1) and (0)). We
take the similar ansatz with two
 unknown functions
 $u_0=u_{-1}$ and $u_1=u_{-2}$.
Now we obtain the two following
 equations for the case $n=0$:
\begin{equation}
0=(C+A-2A)-\omega_d^2(A-{B^2\over2}-A^2)
\label{enegalzeroconstbis}
\end{equation}

\begin{equation}
-\omega_b^2 B=(B+D-2B)-\omega_d^2(B-2AB)
\label{enegalzeroomegbis}
\end{equation}

An analysis similar to that given above,
 can be carried out and we obtain:
\begin{equation}
{D\over B}={-(1+2A\omega_d^2)+
\sqrt{(1+2A\omega_d^2)^2+4}\over 2}\; .
 \label{eDsurBbis}\end{equation}
and
\begin{equation}
{\omega_b^2\over \omega_d^2}={3\over 2\omega_d^2}+1-A-
\sqrt{{1\over 2\omega_d^4}+\left({1\over 2\omega_d^2}+A
\right)^2}\label{eomegaamplapproxbis}\end{equation}

The comparison of the two equations
 (\ref{eomegaamplapprox}) and
 (\ref{eomegaamplapproxbis}),
with the results of the Green's
functions method is shown in
Fig.~\ref{fig:omegaamplitude}(a).
As might be expected at low amplitude,
the results
tends to the NLS case, with a good agreement.
In this domain, as the excitation
concern more than three or four
particles contrary to the postulate in
the ans\" atze (\ref{ansatzun}) and
 (\ref{ansatzdeux}), the present formalism
failed and the agreement is poor.
But, in the highly localized
regime, in which  we are essentially
interested,
 the two expressions given
by the simple anz\"atze are valid.
Although the method can seem very crude,
it provides accurate
results in the very discrete cases
 because the solution are naturally
well localized so that the displacements
which are ignored
 here are really very small.

Using these results we
 can obtain the energy of the mode.
The expression for the case centered
 between particles is
\begin{eqnarray}
E_c={1\over2}u_1^2+(u_1-u_0)^2&&+
\omega_d^2\left({u_0^2\over 2}-
{u_0^3\over3}\right)\nonumber\\
&&+2\omega_d^2\left({u_1^2\over 2}-
{u_1^3\over3}\right)
\end{eqnarray}
\narrowtext
and
\begin{equation}
E_d=E_c+\omega_d^2\left({u_0^2\over 2}-
{u_0^3\over3}\right)
\end{equation}
in the other case, where the expression
of the two displacements
 $u_0$ and $u_1$ are easy determined,
 using $A$, $B$, $C$ and $D$.
The results shown in Fig.~\ref{fig:omegaamplitude}(b),
 attest that, despite
its simplicity, the
calculation gives accurate results
 especially in the second case.

\section{ LOCALIZATION BY COLLISIONS}
\label{sec:Collisions}

To study the interactions between the breathers,
we must rely on numerical
simulations since, in the discrete model, no exact
 solution is available.
In the energy localization process that
 we propose, small amplitude
breathers are generated by spontaneous modulation
 of some energy initially
evenly distributed in the system, and then collisions
favor the growth of
some of the excitations at the expense of the others.
The process requires
generally several collisions.
In order to study this
effect in a controlled manner, we have
 confined two breathers between two
impurity sites where the on-site potential
$V(u)$ is removed. These sites
act as perfectly reflecting walls for
the breathers which bounce back and
forth between the defects. If two
 solitons were sent toward each other in
such a system they would simply pass
through each other many times as they
oscillate in the ``box''. For discrete
 breathers, the picture is very
different. Fig.~\ref{fig:collisions}
shows a typical numerical simulation
result. To generate this figure, two
breathers of unequal amplitude have
been sent toward each other. After 5
collisions, only a large amplitude
breather subsists in the system and the
smaller excitation can no longer be
distinguished from the small amplitude
 waves which have been radiated
during the collisions. Moreover, as one
 of the breathers grows in amplitude,
its PN barrier increases and the breather
 is finally completely trapped by
discreteness. It is important to notice
 however that it is still slowly
growing as shown in fig.~\ref{fig:enggrowth}
 because it collects some energy
of the small amplitude waves generated
in the collision. The detail of the
interaction between discrete breathers
depends on the precise conditions of
the collision, and in particular on the
 relative phases of the two breathers
when they collide. It may even happen
that, in a single collision, the
bigger breather loses some energy.
However, we have observed that the
average effect of multiple collisions
occuring randomly in a lattice,
is always to
increase the amplitude of the larger excitations.
{\it This phenomenon is
very general and very robust to perturbations.}
In particular, the same
behavior is found in a thermalized system, which
 is important for physical
applications. To check this point, we have
 prepared thermalized lattices by
running constrained temperature numerical
simulations with the Nose scheme
\cite{NOSE}.

Then we have launched couples of breathers
 in the chain and
noticed again that the bigger breather grows
at the expense of the smaller one. In fact,
 we observe that
its growth rate is larger in the presence
of thermal fluctuations because it
collects some energy from the fluctuations.
 The results do not depend on the
boundary conditions. Multiple collisions
 can also
be generated by periodic boundary conditions
and the same results are found.
More importantly, the results do not depend
on the particular nonlinear
lattice model which is considered.
Using the more physical Morse potential
instead of $V(u)$ given by Eq.
(\ref{eq:epotentiel}) leads to the same
general conclusions.

\section{CONCLUSION}
\label{sec:CONCLUSION}

Discreteness can be viewed as a perturbation
 of the integrable Nonlinear
Schr\"odinger equation which can be
derived for many nonlinear lattice
models in the continuum and medium amplitude
 limit. Therefore, one might
have expected that the usual property
of the solitons of passing through
each other without energy exchange would be
 destroyed as the integrability
is lost. This is however not so obvious
 because, in the first order of
perturbation, conservative perturbations
do not cause energy exchange in
two-soliton collisions \cite{KIVSHARMALOMED}.
 Moreover, the most remarkable
result is that the world of discrete solitons
is as merciless for the weak
as the real world: in the presence of
discreteness, breather interactions
show a systematic tendency to favor the growth of
 the larger excitation at
the expense of the others.

However, the process contains also its own
regulation mechanism because of the fast increase of the Peierls
barrier with
the amplitude of the breathers.
When they become large enough, the breathers
stay trapped by discreteness. As a result,
 energy initially evenly
distributed over the lattice tends to
concentrate itself into large
amplitude breathers, but the localization
stops before all the energy has
collapsed into a single very large excitation.
 The mechanism of
discreteness-induced energy localization
 that we have described here can
appear in a large variety of physical
 systems involving lattices. In
particular, it is clearly at work
in a model of nonlinear DNA dynamics that
we have investigated recently \cite{DPB}.
 Numerical simulations of the model
at constrained temperature show that,
in the steady state, thermal energy
tends to localize itself around some
sites and consequently the lattice in
equilibrium is very far from equipartition of energy.

\acknowledgements
T.D. and M.P. acknowledge the hospitality
 of the Center for Nonlinear Studies
of the Los Alamos National Laboratory
 where part of this work has been done.
We would like to thank S. Aubry and
A. R. Bishop for useful discussions,
and  Y. S. Kivshar and D. K. Campbell
 for communicating their work
prior to publication.
The authors thank CEC for financial
support with the contract No.
 SC1-CT91-0705.

\begin{figure}
\caption{Profiles of the centered and
non-centered breather solutions
at the time corresponding to the
 maximum amplitude, for a breather
frequency $\omega_b = 0.873~ \omega_d$,
 with $\omega_d^2 = 10$.}
\label{fig:profilsolu}
\end{figure}

\begin{figure}
\caption{  Comparison of the centered
and decentered brea\-ther.
(a) Frequency of the breather modes
 versus amplitude.
The solid line refer to the equation
({\protect \ref{eomegaamplapprox}}),
 the dotted line to the equation
 ({\protect \ref{eomegaamplapproxbis}})
 and the dash-dot-dot-dotted to
the NLS approximation.
(b) Total energy as a function of
 the frequency.
The circles (resp. the plus signs)
 correspond to the
solution obtained with the Green's
 function technic when
 the breather is centered on a particle
 (resp. between two particles).}
\label{fig:omegaamplitude}
\end{figure}

\begin{figure}
\caption{Envelop of the oscillations
of the particle
 24(a), 25(c) and 26(b), when the breather
 is centered between
 the 24 and 25 ones, at the beginning
of the simulations.
After about 300 breather oscillations,
 the excitation moves down to the
Peierls-Nabarro well.}
\label{fig:instabilite}
\end{figure}

\begin{figure}
\caption{Numerical simulation of
 the time evolution of two discrete
breathers sent toward each other
between two reflecting defects situated at sites
30 and 70. The initial amplitudes of the breathers are in the
ratio $A_{\text {right}} / A_{\text {left}} = 1.36$.
The figure shows the energy density in the
discrete chain using a contour plot.
 Darker regions correspond to
regions where the energy density is higher.}
\label{fig:collisions}
\end{figure}

\begin{figure}
\caption{Time evolution of the energy
 of the three central particles of the
biggest breather in the numerical simulation of
fig.~\protect\ref{fig:collisions}. }
\label{fig:enggrowth}
\end{figure}

\end{document}